\documentclass{PoS}
\def\msbar{{\overline{\rm MS}}}
\def\Tr{{\rm Tr}}

\title{Non-perturbative renormalization of overlap quark bilinears on domain wall fermion configurations}

\ShortTitle{Non-perturbative renormalization of overlap quark bilinears on DWF configurations}

\author{\speaker{Zhaofeng Liu}\thanks{partially supported by NSFC under Grant 11105153.}
        , Ying Chen\thanks{partially supported by NSFC under Grants 10835002 and 11075167.}, Yi-Bo Yang\\
        Institute of High Energy Physics and Theoretical Physics Center for Science Facilities, Chinese Academy of Sciences, Beijing 100049, China\\
        E-mail: \email{liuzf@ihep.ac.cn}}

\author{Shao-Jing Dong, Michael Glatzmaier, Ming Gong, Keh-Fei Liu\\
        Department of Physics and Astronomy, University of Kentucky, Lexington, KY 40506
}

\author{Anyi Li\\
        Institute for Nuclear Theory, University of Washington, Seattle, WA 98195
}

\author{Jian-Bo Zhang\thanks{partially supported by NSFC under Grant 11275169.}\\
        Department of Physics, Zhejiang University, Hangzhou 311027, China
}

\author{($\chi$QCD Collaboration)}

\abstract{We present renormalization constants of overlap quark bilinear operators on 2+1-flavor domain wall fermion configurations. 
Both overlap and domain wall fermions have chiral symmetry on the lattice. The scale independent renormalization
constant for the local axial vector current is computed using a Ward Identity. The renormalization constants for the scalar, pseudoscalar and vector current are 
calculated in the RI-MOM scheme. Results in the $\overline{\rm MS}$ scheme are obtained by using perturbative conversion ratios. 
The analysis uses in total six ensembles with lattice sizes $24^3\times64$ and $32^3\times64$.}

\FullConference{31st International Symposium on Lattice Field Theory LATTICE 2013\\
                 July 29 - August 3, 2013\\
                 Mainz, Germany}

\begin{document}

\section{Introduction}
The setup of overlap valence quark on domain wall fermion configurations is successful 
in lattice
calculations of physical quantities such as the strangeness in nucleon~\cite{Gong:2013vja}. The inversion of overlap fermions can be speed up 
by using HYP smearing~\cite{Hasenfratz:2001hp} and 
deflation with low eigenmodes~\cite{Li:2010pw}. The $\chi$QCD collaboration is determining charm and strange quark masses~\cite{ybyang} and 
other physical quantities using this setup. 
The renormalization constants of quark bilinear operators needed to match lattice results to those in the continuum $\overline{\mbox{MS}}$ scheme 
are calculated in this work.

We use the RI-MOM scheme~\cite{Martinelli:1994ty} to calculate renormalization constants for flavor non-singlet scalar, pseudoscalar, vector and axial vector operators
$\mathcal{O}=\bar\psi\Gamma\psi'$, where $\Gamma=I, \gamma_5, \gamma_\mu, \gamma_\mu\gamma_5$ respectively.
The results are converted to the $\msbar$ scheme. 
We have two lattice spacings with
various 
quark masses and give results
in the chiral limit of both the valence and light sea quark masses. 

\section{Methodology}\label{sec:ri_and_overlap}
The RI-MOM scheme~\cite{Martinelli:1994ty} imposes renormalization conditions on amputated Green functions of
the relevant operators in the momentum space. The Green functions needed to be computed include the quark propagator, the forward Green function
and the vertex function:
\begin{equation}
S(p)=\sum_x e^{-ipx}\langle\psi(x)\bar\psi(0)\rangle,\quad
G_{\mathcal{O}}(p)=\sum_{x,y}e^{-ip\cdot
(x-y)}\langle\psi(x)\mathcal{O}(0)\bar{\psi}(y)\rangle,
\end{equation}
\begin{equation}
\Lambda_{\mathcal{O}}(p)=S^{-1}(p)G_{\mathcal{O}}(p)S^{-1}(p).
\end{equation}
The renormalization condition is (imposed in the chiral limit)
\begin{equation}
Z_q^{-1}Z_{\mathcal{O}}\frac{1}{12}
\Tr\left[\Lambda_{\mathcal{O}}(p)\Lambda_{\mathcal{O}}^{tree}(p)^{-1}\right]_{p^2=\mu^2}\equiv Z_q^{-1}Z_{\mathcal{O}}\Gamma_{\mathcal{O}}(p)|_{p^2=\mu^2}=1,
\label{eq:ri_condition}
\end{equation}
where $Z_q$ is the quark field renormalization constant with $\psi_R=Z_q^{1/2}\psi$ (the subscript ``R" means after renormalization), 
$Z_{\mathcal{O}}$ is the renormalization constant for operator $\mathcal{O}$ with
$\mathcal{O}_R=Z_{\mathcal{O}}\mathcal{O}$ and $\Gamma_{\mathcal{O}}(p)$ is the projected vertex function. 
In practice, we do calculations at finite quark masses and then extrapolate to the chiral limit.
The Green functions in Eq.(\ref{eq:ri_condition}) are not gauge invariant, therefore the calculation has to be
done in a fixed gauge, usually in the Landau gauge.

We compute the renormalization constant $Z_A^{WI}$ of the local axial vector current from a Ward Identity (see Sec.~\ref{sec:za}),
which equals to $Z_A^{RI}$ in the RI scheme. 
Then we use
\begin{equation}
Z_q^{RI}=Z_A^{WI}\frac{1}{12}\Tr\left[\Lambda_{A}(p)\Lambda_{A}^{tree}(p)^{-1}\right]_{p^2=\mu^2}.
\label{eq:zq_ri}
\end{equation}
to get $Z_q^{RI}$, and Eq.(\ref{eq:ri_condition}) to compute $Z_S$, $Z_P$ and $Z_V$
for the scalar, pseudoscalar and vector current. At tree level, $\Lambda_{\mathcal{O}}^{tree}(p)=\Gamma$ for quark bilinear operators.

The overlap operator~\cite{Neuberger:1997fp} is defined as
$D_{ov}  (\rho) =   1 + \gamma_5 \varepsilon (\gamma_5 D_{\rm w}(\rho))$,
where $\varepsilon$ is the matrix sign function and $D_{\rm w}(\rho)$ is the usual Wilson fermion operator, 
except with a negative mass parameter $- \rho = 1/2\kappa -4$ in which $\kappa_c < \kappa < 0.25$. 
We set $\kappa = 0.2$ in our calculation that corresponds to $\rho = 1.5$. The massive overlap Dirac operator is defined as
$D_m = \rho D_{ov} (\rho) + m\, (1 - \frac{D_{ov} (\rho)}{2}) = \rho + \frac{m}{2} + (\rho - \frac{m}{2})\, \gamma_5\, \varepsilon (\gamma_5 D_w(\rho))$.
More details on our point source overlap fermion propagators can be found in Ref.~\cite{Li:2010pw}.
With the good chiral properties of overlap fermions,
we should and indeed find $Z_S=Z_P$ and $Z_V=Z_A$ in our results.

\section{Numerical results}
\label{sec:num_results}
We use configurations generated by the RBC-UKQCD collaboration using 2+1 flavor domain wall fermions\cite{Aoki:2010dy,Allton:2008pn}. 
We employ HYP smearing on the gauge fields and then fix to Landau gauge.
The parameters and statistics of configurations 
are collected in Tab.~\ref{tab:nconfs}.
\begin{table}
\begin{center}
\caption{The parameters and statistics of configurations used in this work. The residual masses in lattice units $m_{res}$
are in the two-flavor chiral limit as given in Ref.~\cite{Aoki:2010dy}.}
\begin{tabular}{cccccc}
\hline\hline
$1/a$ (GeV) & label & $m_l/m_s$ & volume & $N_{conf}$ & $m_{res}$ \\
\hline
1.73(3) & c005  & 0.005/0.04 & $24^3\times64$ & 92 & 0.003152(43) \\
& c01  & 0.01/0.04 & $24^3\times64$ & 88  & \\
& c02 & 0.02/0.04 & $24^3\times64$ & 138 & \\
\hline
2.28(3) & f004 & 0.004/0.03 & $32^3\times64$ & 50 & 0.0006664(76) \\
& f006 & 0.006/0.03 & $32^3\times64$ & 40 & \\
& f008 & 0.008/0.03 & $32^3\times64$ & 50 & \\
\hline\hline
\end{tabular}
\label{tab:nconfs}
\end{center}
\end{table}
The overlap valence quark masses are given in Tab.~\ref{tab:24_32}. 
The corresponding pion masses are from about 220 to 600 MeV.
\begin{table}
\begin{center}
\caption{Overlap valence quark masses in lattice units on the $24^3\times64$ and $32^3\times64$ lattices.}
\begin{tabular}{ccccccccc}
\hline\hline
$24^3\times64$  & 0.00620 & 0.00809 & 0.01020 & 0.01350 & 0.01720 & 0.02430 & 0.03650 & 0.04890 \\
\hline
$32^3\times64$ &   0.00460 & 0.00585 & 0.00677 & 0.00885  &   0.01290 & 0.01800 & 0.02400 & 0.03600 \\ 
\hline\hline
\end{tabular}
\label{tab:24_32}
\end{center}
\end{table}

We use (anti-)periodic boundary condition in the spacial(time) directions. Thus the momenta are
$ap=(\frac{(2k_t+1)\pi}{T}, \frac{2\pi k_x}{L}, \frac{2\pi k_y}{L}, \frac{2\pi k_z}{L})$,
where $k_\mu=-6,-5,...,6$ on the $L=24$ lattice and $k_t=-5,-1,...,6$, $k_i=-6,-5,...,6$ on the $L=32$ lattice. To reduce the effects of Lorentz
non-invariant discretization errors, we use the momenta which satisfy the condition
$\frac{p^{[4]}}{(p^2)^2}<0.32, \mbox{where } p^{[4]}=\sum_\mu p_\mu^4\mbox{ and }p^2=\sum_\mu p_\mu^2$.
The statistical errors of our final results are from Jackknife processes.

\subsection{Renormalization of the local axial vector current}
\label{sec:za}
The renormalization constant $Z_A$ can be obtained from the axial Ward identity
\begin{equation}
Z_A\partial_\mu A_\mu=2Z_m m_q Z_P P,
\label{eq:ZA_WI}
\end{equation}
where $A_\mu$, $P$ are the local axial vector current and the pseudoscalar density and $Z_m$ is the quark mass renormalization constant with $m_R=Z_m m_q$.
Since $Z_m=Z_P^{-1}$ for overlap fermions,
one can find $Z_A$ by considering the matrix elements of the both sides of Eq.(\ref{eq:ZA_WI}) between the vacuum and a pion:
$Z_A\partial_\mu\langle 0| A_\mu |\pi\rangle=2m_q\langle 0|P|\pi\rangle$.
If the pion is at rest, one has
$Z_A=\frac{2m_q\langle0|P|\pi\rangle}{m_\pi\langle 0| A_4 |\pi\rangle}$,
where $A_4=\bar\psi\gamma_4\gamma_5\psi$.
To obtain the matrix elements, we compute 2-point correlators
$G_{PP}(\vec p=0,t)=\sum_{\vec x}\langle0|P(x)P(0)|0\rangle\mbox{ and }G_{A_4P}(\vec p=0,t)=\sum_{\vec x}\langle0|A_4(x)P(0)|0\rangle$.
Here we follow Ref.~\cite{Zhang:2005sca} closely.
When the time $t$ is big, the contribution from the pion dominates in both correlators. Then one has
\begin{equation}
Z_A^{WI}=\lim_{m_q\rightarrow0,t\rightarrow\infty}\frac{2m_qG_{PP}(\vec p=0,t)}{m_\pi G_{A_4P}(\vec p=0,t)}.
\label{eq:Za_WI_2pt}
\end{equation}

\begin{figure}[t!]
\null
\vspace{-4ex}
\begin{minipage}[t]{.48\linewidth}
\centerline{\includegraphics[width=0.6\linewidth]{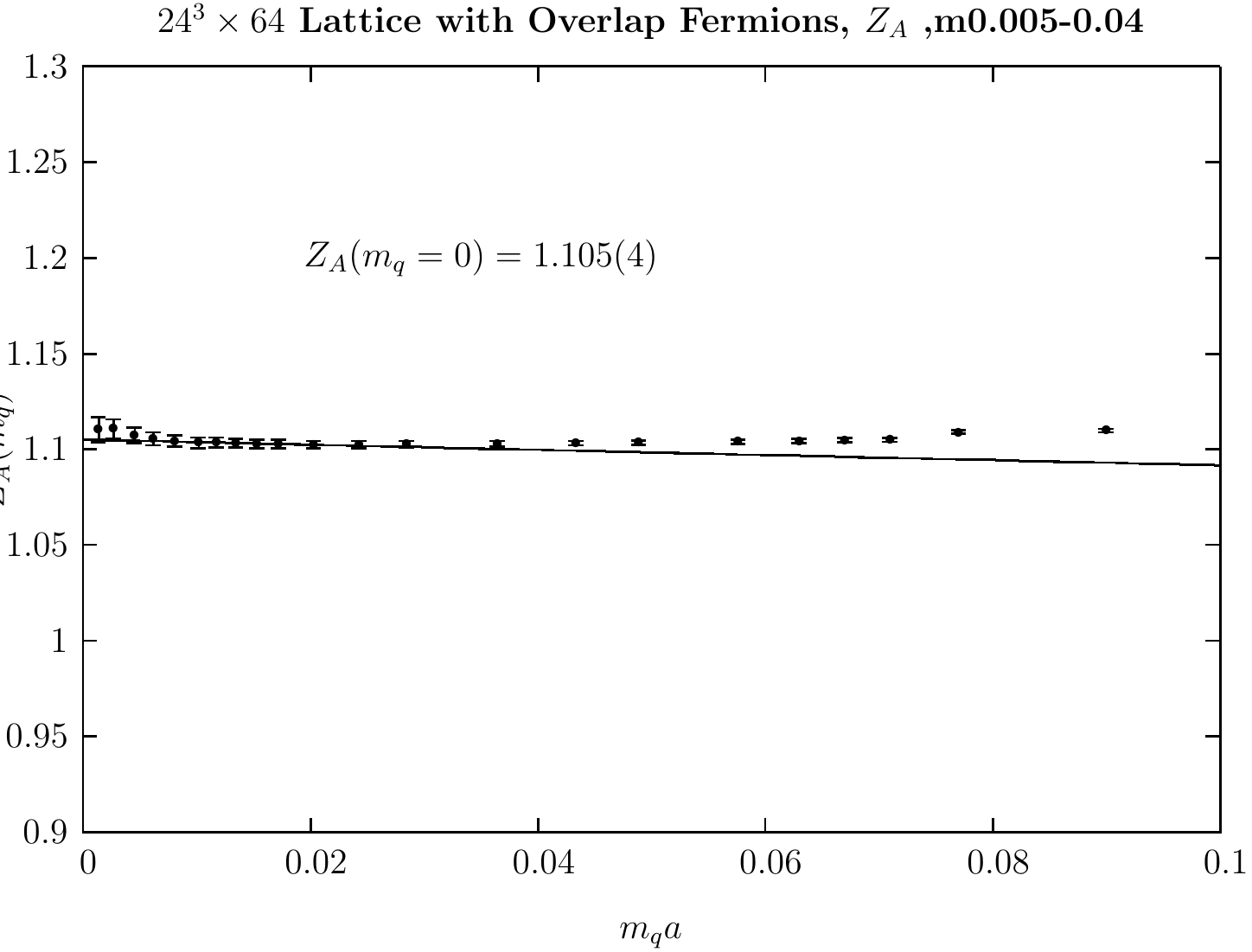}}
\caption{$Z_A^{WI}$ against valence quark masses on data ensemble c005.}
\label{fig:Za_WI}
\end{minipage}
\hfill
\begin{minipage}[t]{.48\linewidth}
\centerline{\includegraphics[width=0.6\linewidth,height=1.3in]{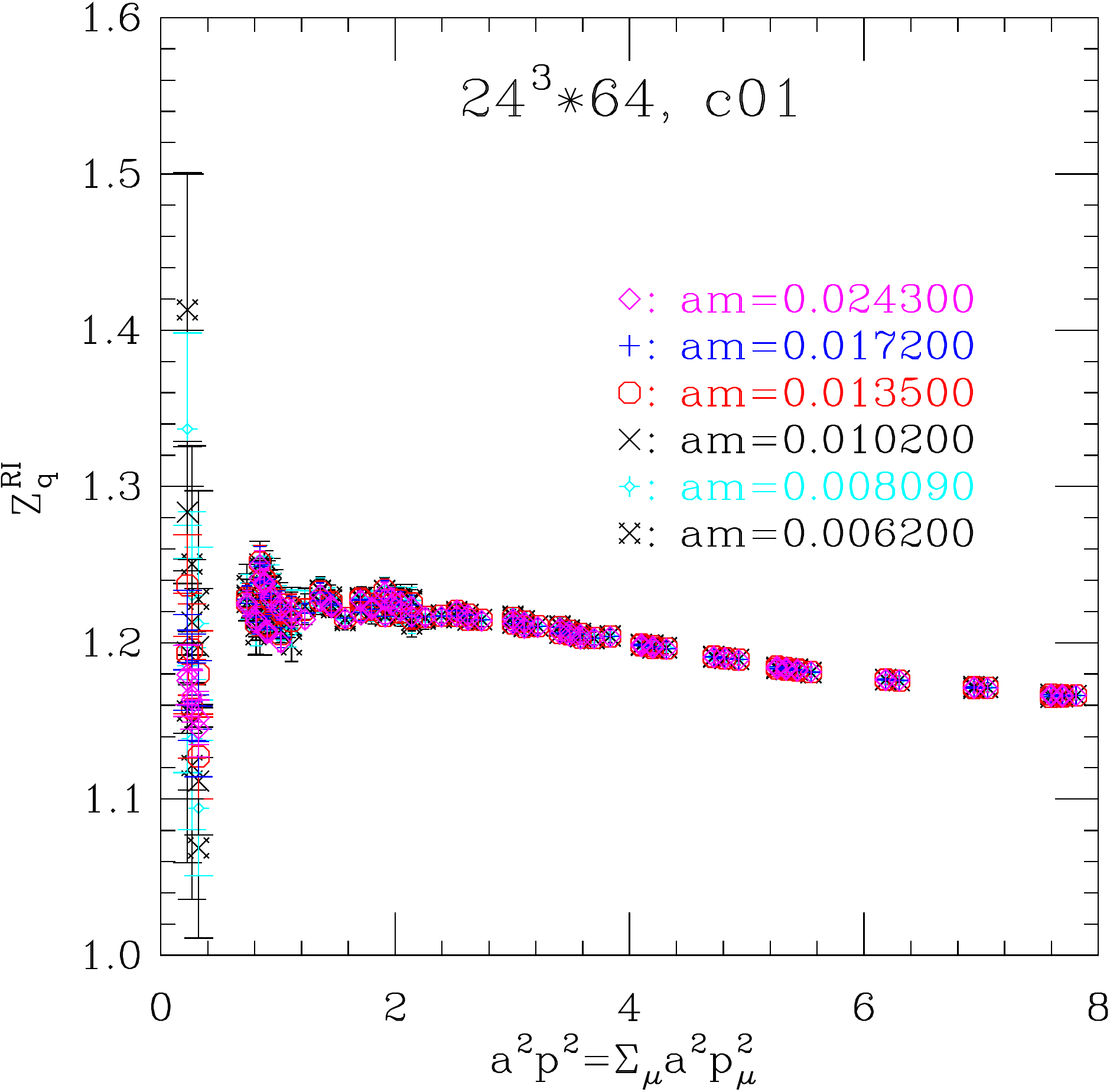}}
\caption{$Z_q^{RI}$ as a function of the momentum scale for different valence quark masses on ensemble c01.}
\label{fig:zq_ri}
\end{minipage}
\end{figure}

Fig.~\ref{fig:Za_WI} shows an example of $Z_A^{WI}$ obtained from Eq.(\ref{eq:Za_WI_2pt}) before taking the quark massless limit.
For the valence quark massless limit, we fit the data to
$ Z_A^{WI}=\hat Z_A^{WI}(1+b_A am_q + c_A (am_q)^2)$.
The results of $\hat Z_A^{WI}$ are given in Tab.~\ref{tab:Za_WI_24_32}.
\begin{table}
\begin{center}
\caption{$\hat Z_A^{WI}$ on the $24^3\times64$ and $32^3\times64$ lattices.}
\begin{tabular}{cccccc}
\hline\hline
$24^3\times64$ & $m_l/m_s$ & 0.005/0.04 & 0.01/0.04 & 0.02/0.04 & $m_l+m_{res}=0$ \\
 & $\hat Z_A^{WI}$ & 1.105(4) & 1.115(6) & 1.101(4) & 1.111(6) \\
\hline
$32^3\times64$ & $m_l/m_s$ & 0.004/0.03 & 0.006/0.03 & 0.008/0.03 & $m_l+m_{res}=0$ \\
 & $\hat Z_A^{WI}$ & 1.080(1) & 1.079(1) & 1.075(1) & 1.086(2) \\
\hline\hline
\end{tabular}
\label{tab:Za_WI_24_32}
\end{center}
\end{table}
In the last column, the results at the light sea quark massless limit are from a linear extrapolation in $m_l+m_{res}$ with $m_{res}$ given in Tab.~\ref{tab:nconfs}.

\subsection{Renormalization constants of the quark field and local vector current}
Fig.~\ref{fig:zq_ri} shows examples of $Z_q^{RI}$ computed from Eq.(\ref{eq:zq_ri}) against the scale for various
valence quark masses. The quark mass dependence of $Z_q^{RI}$ is quite small.
In Landau gauge, the anomalous dimension of $Z_q$ is zero at 1-loop. This is why $Z_q$ is quite flat in Fig.~\ref{fig:zq_ri}.

\begin{figure}[t!]
\null
\vspace{-4ex}
\begin{minipage}[t]{.48\linewidth}
\centerline{\includegraphics[width=0.6\linewidth,height=1.25in]{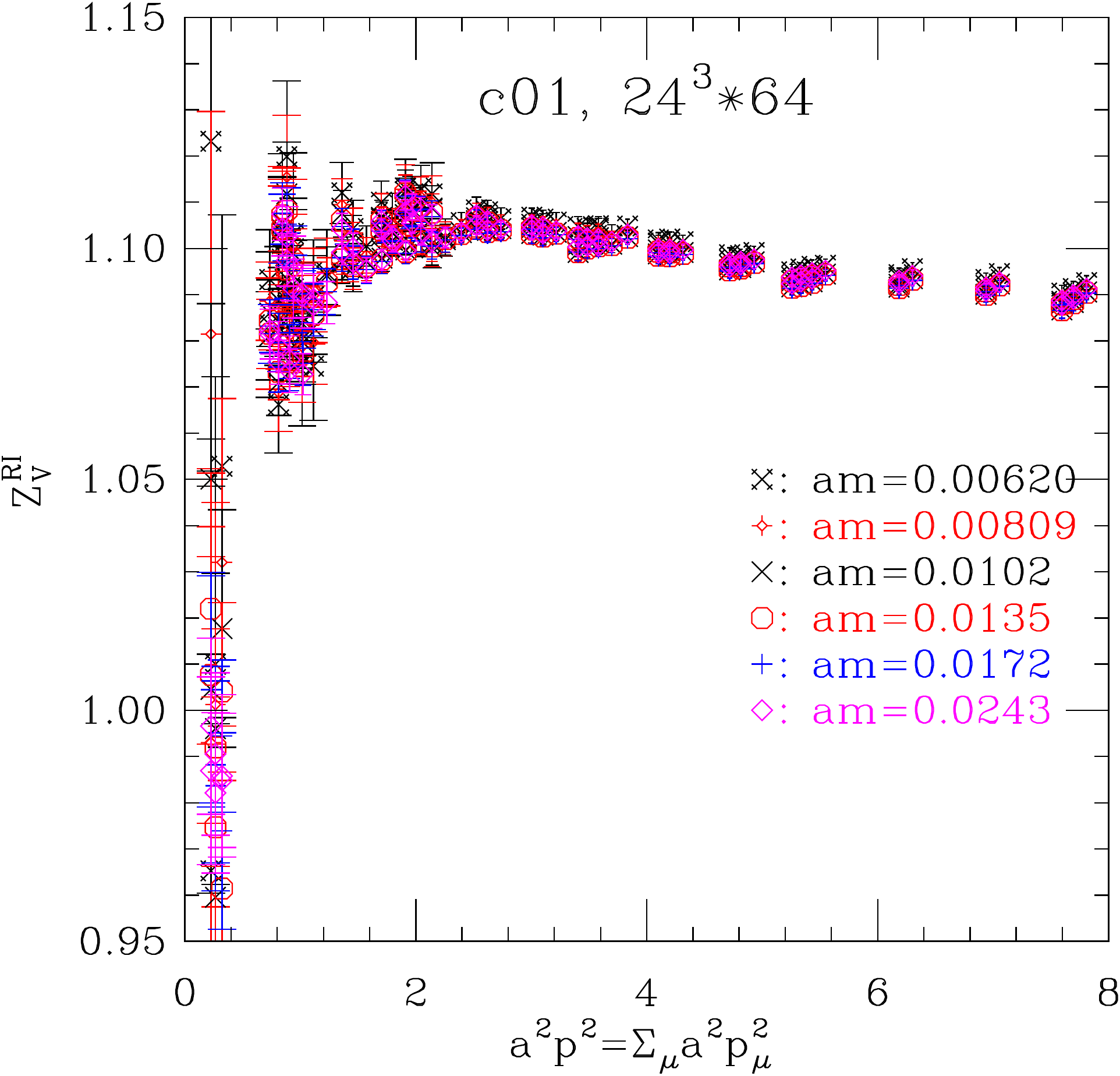}}
\caption{Examples of $Z_V^{RI}$ as functions of the momentum scale for ensemble c01.}
\label{fig:zv_ri}
\end{minipage}
\hfill
\begin{minipage}[t]{.48\linewidth}
\centerline{\includegraphics[width=0.6\linewidth,height=1.25in]{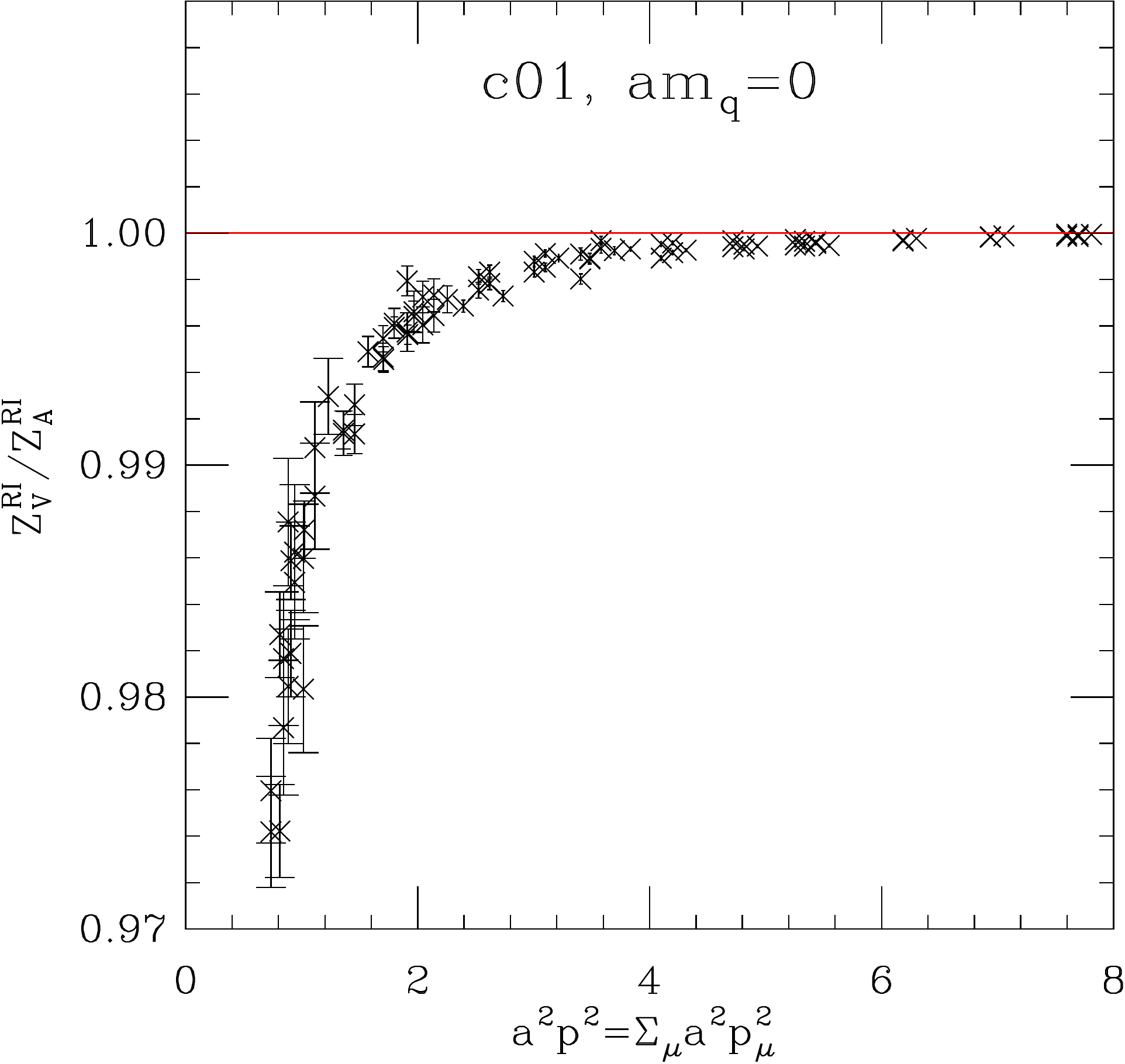}}
\caption{$Z_V^{RI}/Z_A^{RI}$ in the valence quark massless limit against the scale for ensembles c01.}
\label{fig:zv_over_za_ri}
\end{minipage}
\end{figure}

$Z_V^{RI}$ for the local vector current for different valence quark masses on ensemble c01
are shown in Fig.~\ref{fig:zv_ri}. Here in using Eq.(\ref{eq:ri_condition}), we have averaged $\mu=1,2,3,4$ for the vector current.
The quark mass dependence for $Z_V^{RI}$ is small. $Z_V^{RI}$ is scale independent at large scale:
When $a^2p^2>\sim3$, $Z_V^{RI}$ is flat up to discretization errors.
Fig.~\ref{fig:zv_over_za_ri} shows the ratio $Z_V^{RI}/Z_A^{RI}$ for ensemble c01.
To go to the chiral limit, we used a linear extrapolation in valence quark masses for $Z_V^{RI}/Z_A^{RI}$.
As we see in Fig.~\ref{fig:zv_over_za_ri}, at large momentum scale $Z_V^{RI}/Z_A^{RI}=1$, i.e., $Z_V^{RI}=Z_A^{RI}$ is satisfied as expected.
The results of $Z_V^{RI}/Z_A^{RI}$ for other five ensembles are similar.

\subsection{Renormalizations of the scalar and pseudoscalar density}
\label{sec:zs_ri}
\begin{figure}[b!]
\null
\vspace{-4ex}
\begin{minipage}[t]{.48\linewidth}
\centerline{\includegraphics[width=0.6\linewidth,height=1.3in]{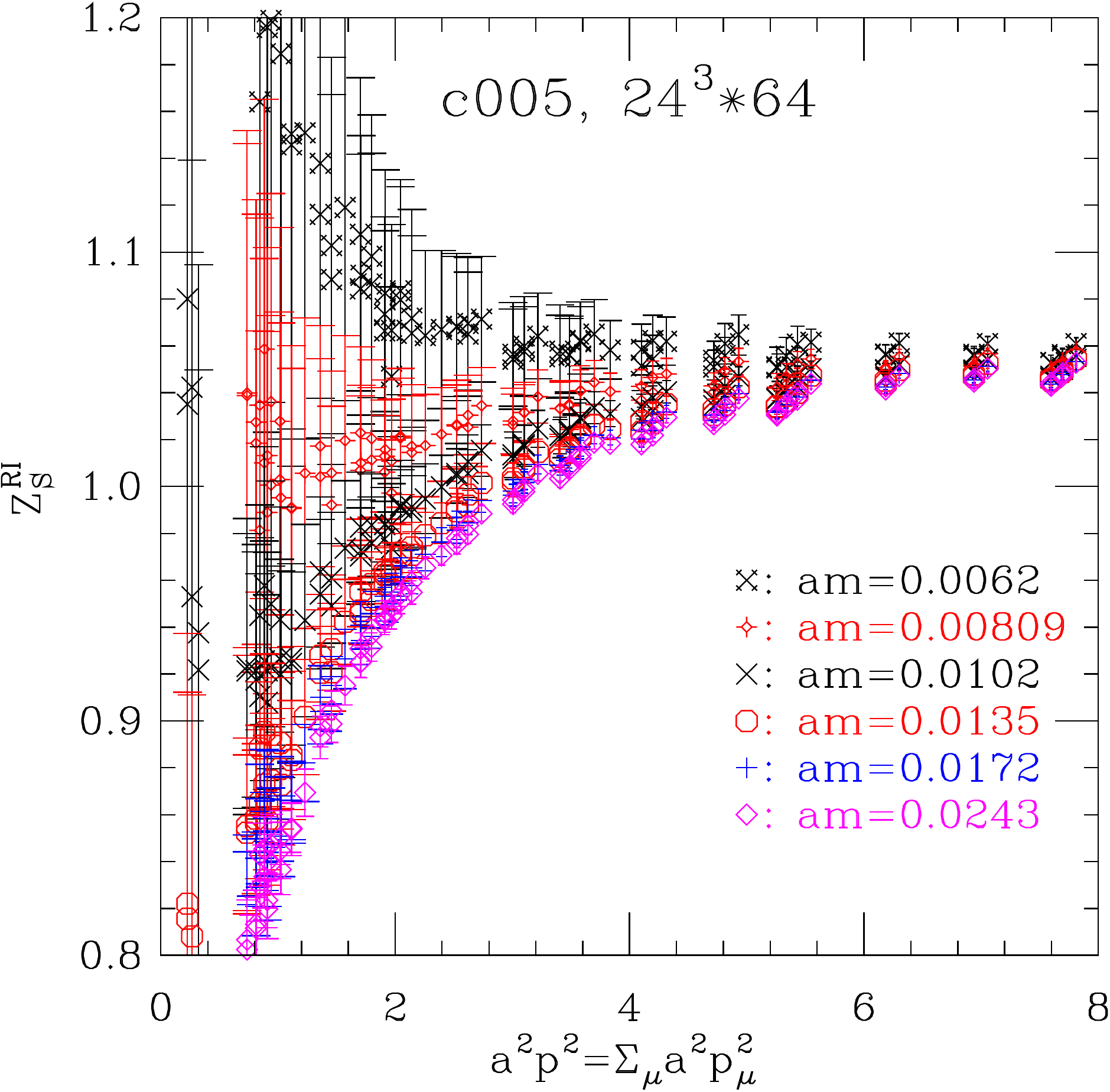}}
\caption{Examples of $Z_S^{RI}$ as functions of the momentum scale for ensemble c005.}
\label{fig:zs_ri}
\end{minipage}
\hfill
\begin{minipage}[t]{.48\linewidth}
\centerline{\includegraphics[width=0.6\linewidth,height=1.3in]{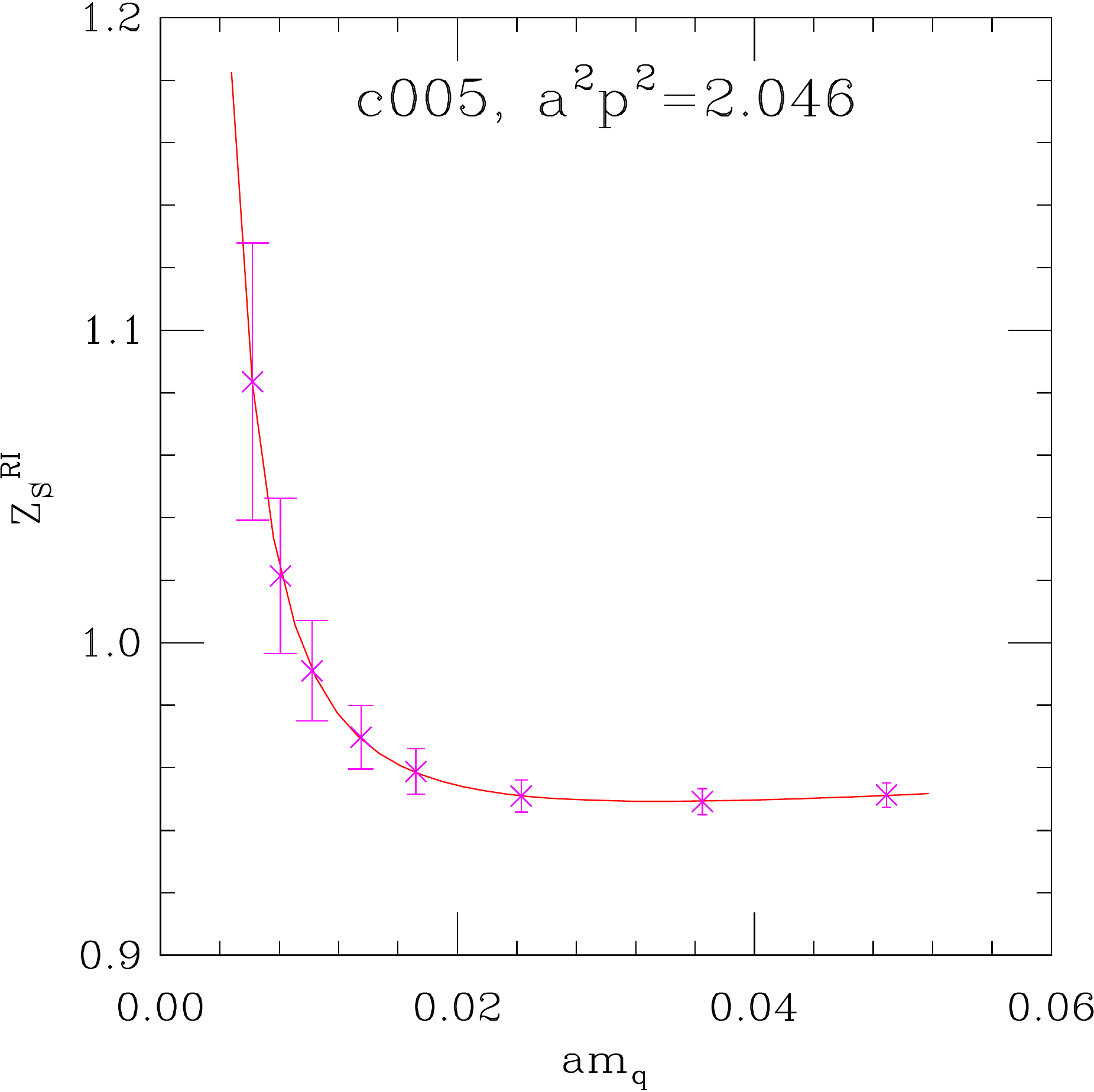}}
\caption{$Z_S^{RI}$ against the valence quark mass at a given scale for ensemble c005.}
\label{fig:zs_amq}
\end{minipage}
\end{figure}

$Z_S^{RI}$ as a function of the scale 
for different valence quark masses ($am_q$) on ensemble c005 are shown in Fig.~\ref{fig:zs_ri}.
Fig.~\ref{fig:zs_amq} shows $Z_S^{RI}$ against $am_q$ at a given scale on ensemble c005.
The dependence on $am_q$ is not linear. To go to the chiral limit, we use a 3-parameter function
\begin{equation}
Z_S=\frac{A_s}{(am_q)^2}+B_s+C_s(am_q)
\label{eq:zs_fit}
\end{equation}
to fit our data and take $B_s$ as the chiral limit value of $Z_S$.
The double pole term comes from topological zero modes~\cite{Blum:2001sr,Aoki:2007xm}.
In a calculation of $Z_S$ in the RI' scheme~\cite{DeGrand:2005af},
the curving up of $Z_S$ at small $am_q$
is suppressed when the zero modes are subtracted from the quark propagator.
\begin{figure}[t!]
\null
\vspace{-4ex}
\begin{minipage}[t]{.48\linewidth}
\centerline{\includegraphics[width=0.6\linewidth,height=1.3in]{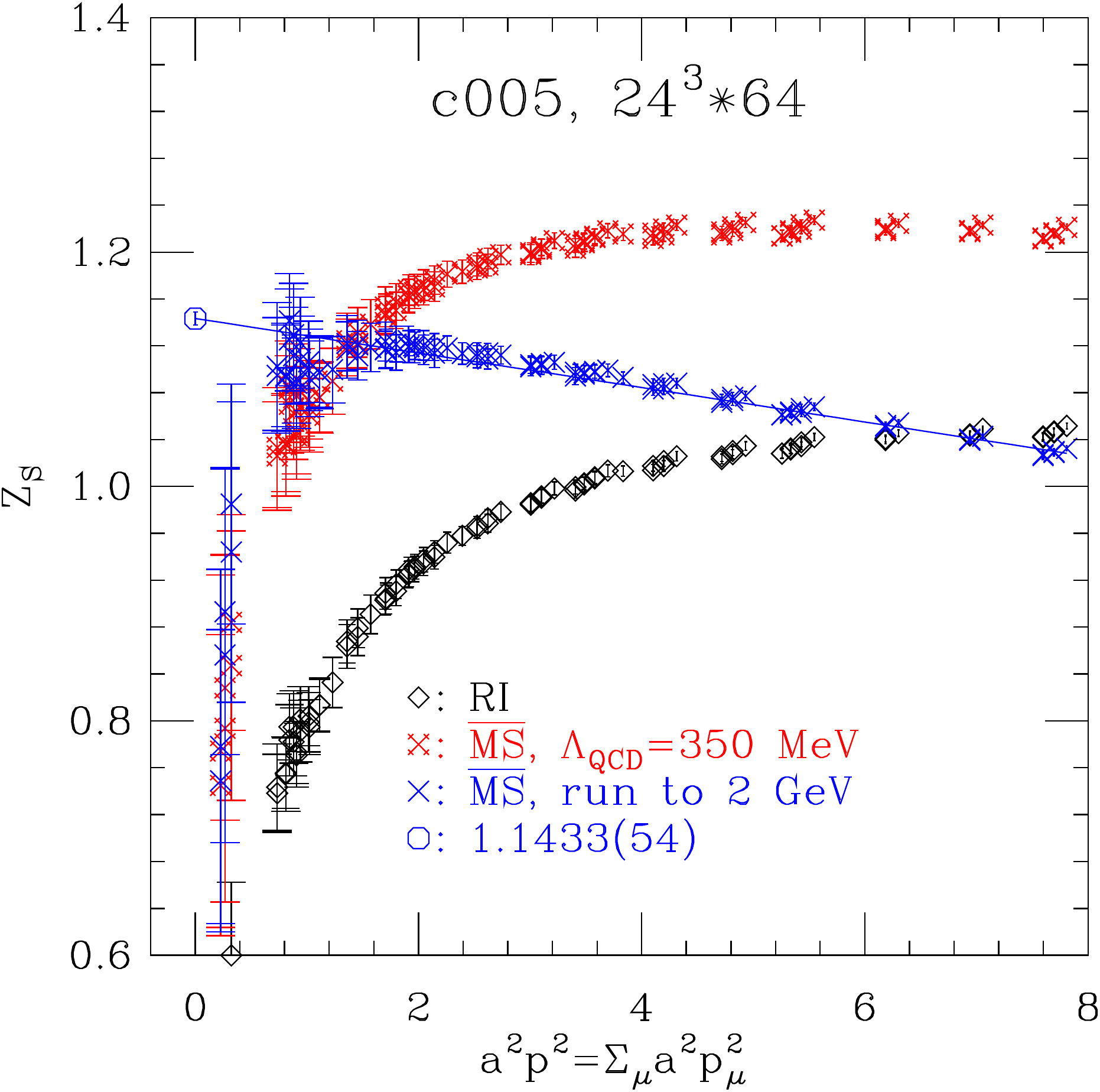}}
\caption{The conversion and running of $Z_S$ in the valence quark massless limit on ensemble c005.}
\label{fig:zs_msbar}
\end{minipage}
\hfill
\begin{minipage}[t]{.48\linewidth}
\centerline{\includegraphics[width=0.6\linewidth,height=1.3in]{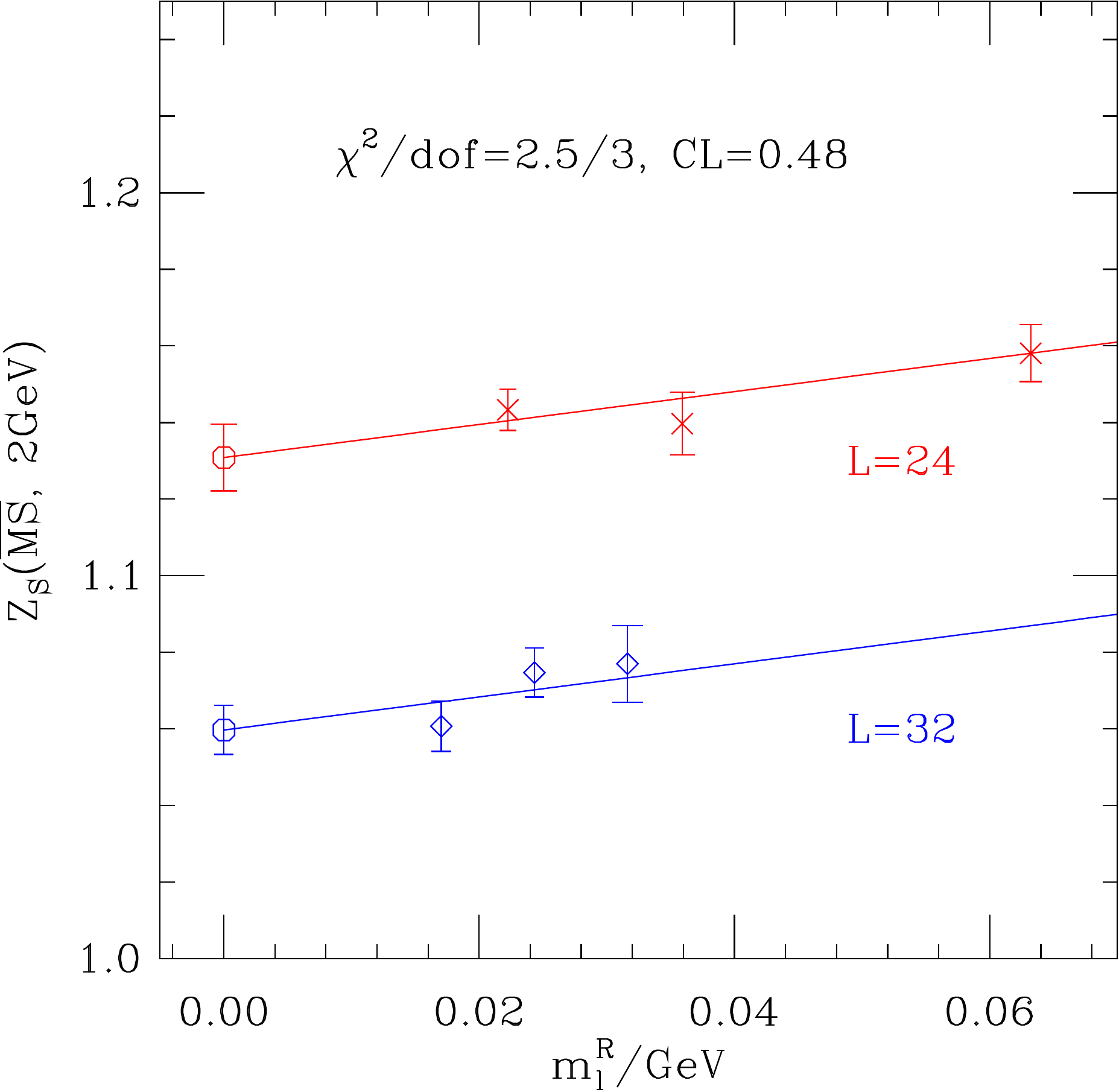}}
\caption{Linear extrapolation of $Z_S^{\msbar}$ to the light sea quark massless limit.}
\label{fig:zs_extra_sea}
\end{minipage}
\end{figure}

The fits to Eq.(\ref{eq:zs_fit}) have small $\chi^2/\mbox{dof}$ at all momentum scales. One example of the fits is shown in Fig.~\ref{fig:zs_amq}.
$Z_S^{RI}$ in the valence chiral limit as a function of the scale for ensemble c005 are shown 
by the black diamonds in Fig.~\ref{fig:zs_msbar}.
We use conversion ratios from continuum perturbation theory~\cite{Chetyrkin:1999pq} to 3-loops to convert $Z_S^{RI}$ to the $\msbar$ scheme.
The value of $\alpha_s^\msbar(\mu)$ in the ratios is obtained by using its perturbative running to 4-loops~\cite{Alekseev:2002zn,vanRitbergen:1997va}.
We use $\Lambda_{QCD}^\msbar=350$ MeV for three flavors.
$Z_S^{\msbar}$ as a function of the scale $a^2p^2$ are shown by the red fancy crosses in Fig.~\ref{fig:zs_msbar}.

To get $Z_S^{\msbar}(2$ GeV$)$, we first use the mass anomalous dimension to 4-loops~\cite{Chetyrkin:1999pq} to
evolve $Z_S^\msbar(a^2p^2)$ to 2 GeV (lattice spacings in Tab.~\ref{tab:nconfs} are used).
The blue crosses in Fig.~\ref{fig:zs_msbar} show $Z_S^\msbar(2$ GeV$; a^2p^2)$, the running results from the
initial scale to 2 GeV. $Z_S^\msbar(2$ GeV$; a^2p^2)$ would lie on a horizontal line at large
$a^2p^2$ if there were no discretization errors.
The solid blue line in Fig.~\ref{fig:zs_msbar} is a linear fit to the blue crosses with $a^2p^2>5$ to extrapolate away 
$\mathcal{O}(a^2p^2)$ discretization errors.
Using the data with $a^2p^2>4$ to do the extrapolation gives consistent results.
$Z_S^\msbar (2$ GeV$)$ on all ensembles are collected in Tab.~\ref{tab:zs}, where we have used
$a^2p^2>5$ for the extrapolations on the $L=24$ lattices and $a^2p^2>3$ on the $L=32$ lattices.
\begin{table}
\begin{center}
\caption{$Z_S^\msbar(2$ GeV) and $Z_{P,\msbar}^{sub}(2$ GeV) on the $24^3\times64$ and $32^3\times64$ lattices.}
\begin{tabular}{ccccc}
\hline\hline
ensemble & c005 & c01 & c02 & $m_{l}+m_{res}=0$  \\
$Z_S^\msbar(2$ GeV) & 1.1433(54) & 1.1397(82) & 1.1581(74) & 1.1308(87) \\
$Z_{P,\msbar}^{sub}(2$ GeV) & 1.164(14) & 1.168(22) & 1.194(28) & 1.141(25) \\
\hline
ensemble &  f004 & f006 & f008 & $m_{l}+m_{res}=0$ \\
$Z_S^\msbar(2$ GeV) &  1.0607(66) & 1.0747(64) & 1.077(10) & 1.0597(64) \\
$Z_{P,\msbar}^{sub}(2$ GeV) & 1.068(21) & 1.093(19) & 1.105(24) & 1.066(21) \\
\hline\hline
\end{tabular}
\label{tab:zs}
\end{center}
\end{table}

From the values on all six ensembles, 
we do a simultaneous linear extrapolation in the 
renormalized light sea quark mass 
to obtain $Z_S^\msbar$ in the sea quark massless limit. The fit function is
\begin{equation}
Z(m_l^R)=Z(0)+c\cdot m_l^R,\quad\mbox{where }m_l^R=(m_l+m_{res})Z_m^{sea}.
\label{eq:extrap_msea}
\end{equation}
Here $Z_m^{sea}=1.578(2)$ on the $L=24$ lattice and $1.573(2)$ on the $L=32$ lattice~\cite{Aoki:2010dy}. The slopes in Eq.(\ref{eq:extrap_msea}) for
the coarse and fine lattices are required to be the same.
The extrapolation is shown in Fig.~\ref{fig:zs_extra_sea}. 
The results after extrapolation are given in the last column of Tab.~\ref{tab:zs}.
We also did separate linear extrapolations in sea quark masses on the coarse and fine lattices and got consistent results. 

The systematic errors of $Z_S$ are given in Tab.~\ref{tab:zs_error}.
\begin{table}
\begin{center}
\caption{Systematic uncertainties of $Z_S^\msbar$(2 GeV) and $Z_{P,\msbar}^{sub}$(2 GeV)}
\begin{tabular}{lcccc}
\hline\hline
 & $Z_S^\msbar$ (L=24) & $Z_S^\msbar$ (L=32) & $Z_{P,\msbar}^{sub}$ (L=24) & $Z_{P,\msbar}^{sub}$ (L=32) \\
Source & Error (\%) & Error (\%)  & Error (\%) & Error (\%) \\
\hline
Truncation (RI to $\msbar$)  &  1.4 & 1.3 &  1.4 & 1.3 \\
Coupling constant  & 1.5 & 1.5 & 1.5 & 1.5 \\
Lattice spacing & 0.5 & 0.4 & 0.5 & 0.4 \\
Fit range of $a^2p^2$ & 0.4 & $<$0.1 & 0.18 & 0.09 \\
Extrapolation in $m_l^R$ & 0.18 & 1.8 & 0.5 & 3.8 \\
Total & 2.2 & 2.7 & 2.2 & 4.3 \\
\hline\hline
\end{tabular}
\label{tab:zs_error}
\end{center}
\end{table}
The $\mathcal{O}(\alpha_s^3)$ term in the conversion ratio from the
RI-MOM to the $\msbar$ scheme contributes about 2.2\%. The ignored $\mathcal{O}(\alpha_s^4)$ term is suppressed by a factor of $\alpha_s$.
Assuming its coefficient is 3 times as that for the $\alpha_s^3$ term, we get a $\sim1.4\%$ truncation error.
Using $\Lambda_{QCD}^\msbar=400$ MeV instead of 350 MeV to evaluate $\alpha_s$ changes $Z_S^\msbar(2$ GeV$)$ by 1.5\% on both lattices.
The $\mathcal{O}(\alpha_s^4)$ term in
the perturbative running of $Z_S^\msbar$ from an initial scale to 2 GeV contributes
less than 0.02\%. Thus this truncation error can be ignored.
Varying the lattice spacings by one sigma leads to $\sim0.5$\% change in
$Z_S^\msbar(2$ GeV$)$.
In the extrapolation of $Z_S^\msbar(2$ GeV;$ a^2p^2)$ to reduce 
discretization errors, changing the range of $a^2p^2$ introduces 0.4\% error on
the coarse lattice or $<0.1$\% error on the fine lattice. Here we vary $a^2p^2>5$ to $>4$ on the coarse lattice 
and $a^2p^2>3$ to $>2$ on the fine lattice.

Finally we get $Z_S^\msbar(2$ GeV)=1.131(9)(25) on the coarse lattice and 1.060(6)(29) on the fine lattice, where the first error
is statistical and the second systematic. 
The statistical uncertainty is
much smaller than the systematic one.

\begin{figure}[t!]
\null
\vspace{-4ex}
\begin{minipage}[t]{.48\linewidth}
\centerline{\includegraphics[width=0.6\linewidth,height=1.22in]{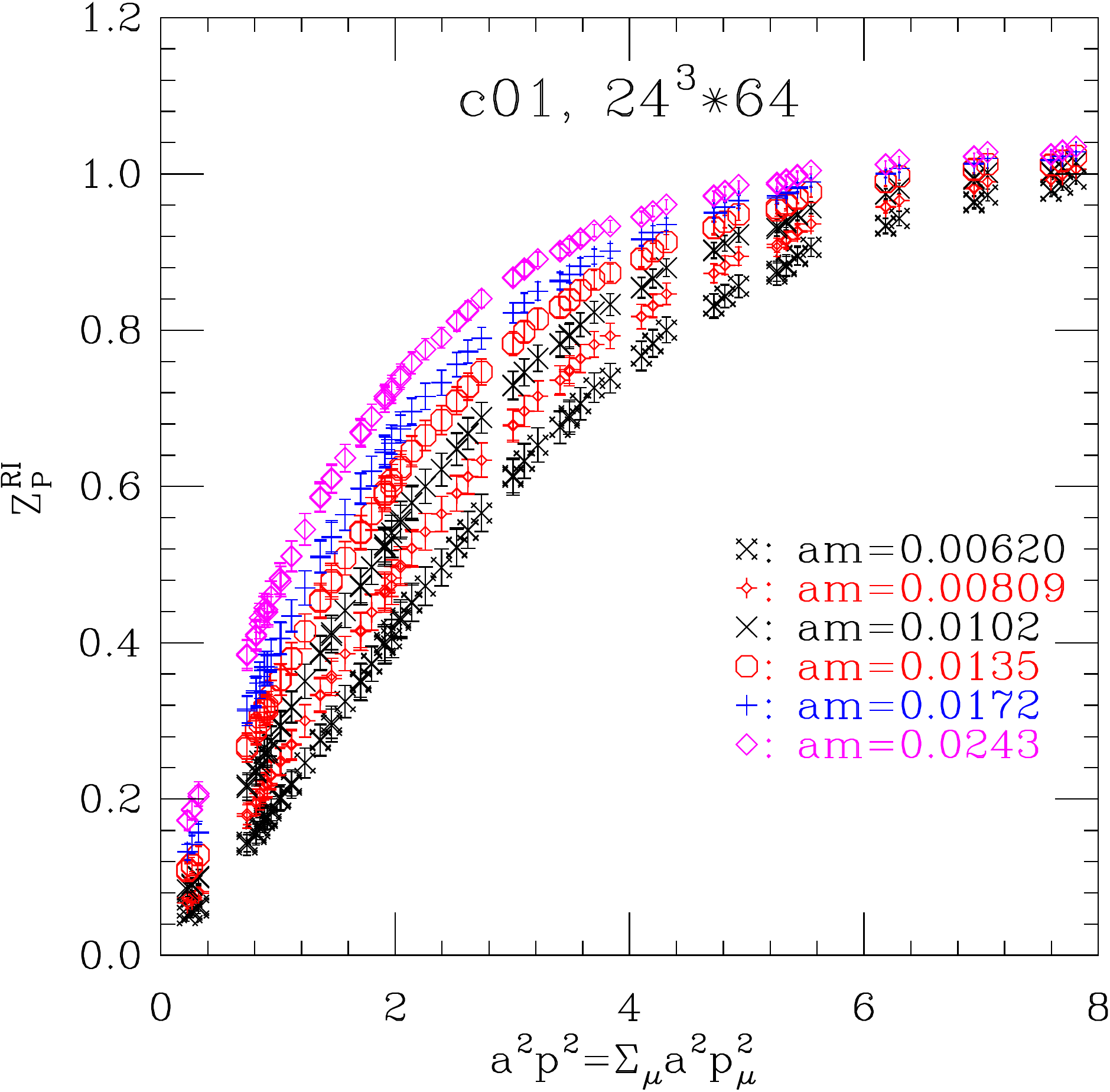}}
\caption{$Z_P^{RI}$ against the momentum scale for ensemble c01 at various valence quark masses.}
\label{fig:zp_ri}
\end{minipage}
\hfill
\begin{minipage}[t]{.48\linewidth}
\centerline{\includegraphics[width=0.6\linewidth,height=1.22in]{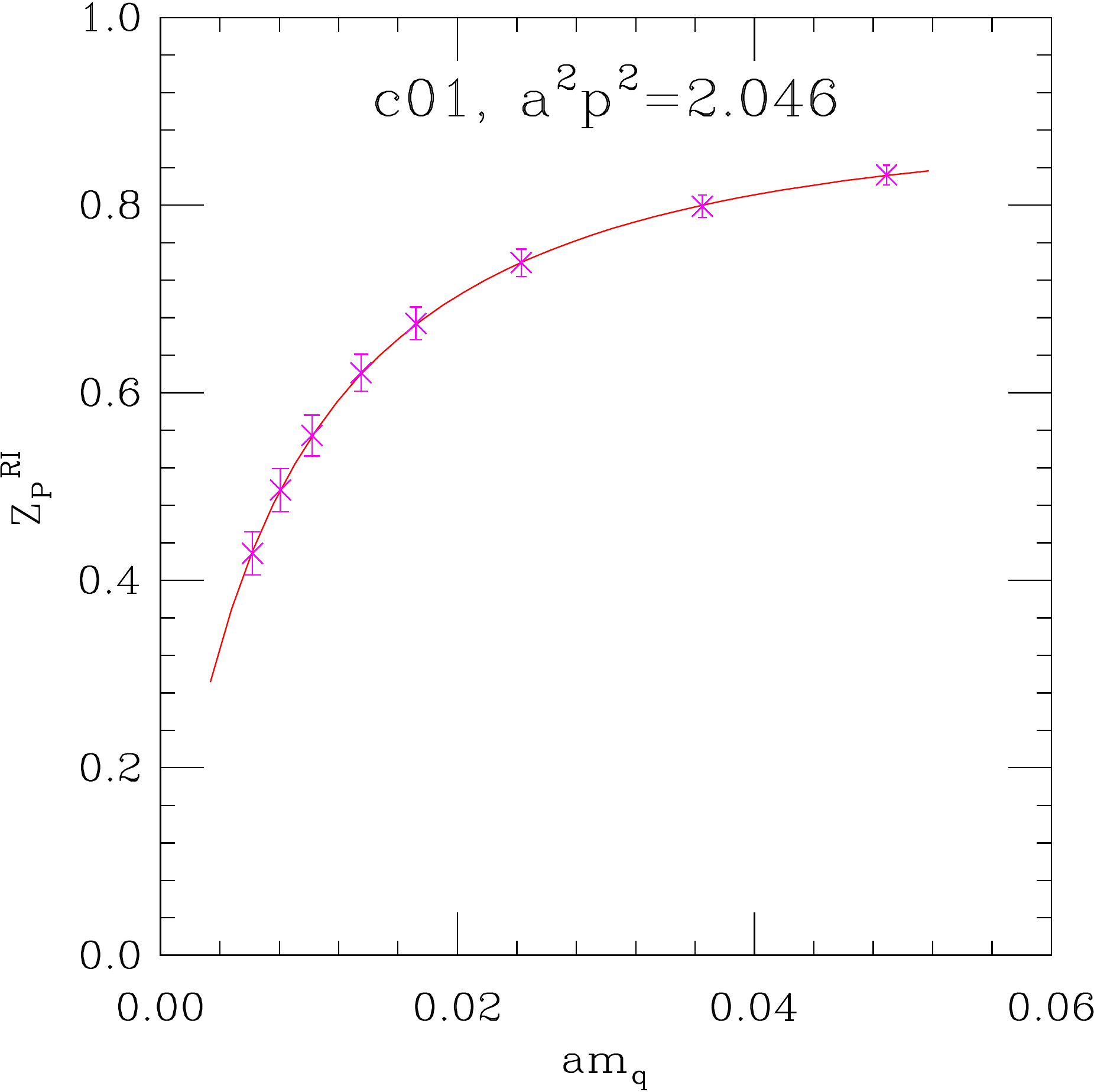}}
\caption{An example of fittings of $Z_P^{RI}$ for ensemble c01.}
\label{fig:zp_extrap_eg}
\end{minipage}
\end{figure}

The pseudoscalar renormalization constant $Z_P^{RI}$ 
is shown in Fig.~\ref{fig:zp_ri} for ensemble c01.
The coupling to the Goldstone boson channel~\cite{Martinelli:1994ty} 
leads to the singular behavior in $Z_P^{RI}$ at small $a^2p^2$.
To remove this non-perturbative effect, we fit $Z_P^{RI}$ at each given $a^2p^2$ to the 3-parameter ansatz~\cite{Becirevic:2004ny}
\begin{equation}
Z_P^{-1}=\frac{A}{am_q}+B+C(am_q),
\label{eq:zp_chiral_ansatz}
\end{equation}
and take $Z_P^{sub}=B^{-1}$ as the RI-MOM value in the valence quark chiral limit.
Fig.~\ref{fig:zp_extrap_eg} shows one example of the fittings of $Z_P^{RI}$ to Eq.(\ref{eq:zp_chiral_ansatz}) 
at a given $a^2p^2$. All the fittings have small $\chi^2/\mbox{dof}$.
$Z_P^{sub}$ is then converted to the $\msbar$ scheme.
Similar to the analysis of $Z_S$, we evolve $Z_{P,\msbar}^{sub}(a^2p^2)$ to 2 GeV.
Then a linear fit in $a^2p^2$ to the data at large $a^2p^2$ is used to reduce $\mathcal{O}(a^2p^2)$ discretization errors. 

$Z_{P,\msbar}^{sub}(2$ GeV$)$ on all ensembles are given in Tab.~\ref{tab:zs}. 
In the last column, the sea quark massless limit values 
are 
obtained from a
simultaneous linear extrapolation in the renormalized light sea quark mass 
with the fit function Eq.(\ref{eq:extrap_msea}).
Comparing the numbers in Tab.~\ref{tab:zs}, we see that $Z_S=Z_P^{sub}$ is well satisfied.
Similar to the analysis for $Z_S$, we give the systematic errors of $Z_P^{sub}$ in Tab.~\ref{tab:zs_error}.
Unlike $Z_S$, the statistical error of $Z_{P,\msbar}^{sub}$ is about the same size as the systematic one.

\end{document}